# Secure quantum routing

Shuo Sun and Edo Waks[a)]

Department of Electrical and Computer Engineering, Institute for Research in Electronics and Applied Physics, and Joint Quantum Institute, University of Maryland, College Park, Maryland 20742, USA

**Abstract**

We show that a quantum network can protect the identity of a sender and receiver from an external wiretapper. This new quantum communication protocol, which we call secure quantum routing, requires only single photons routed by linear optical elements and photon counters, and does not require distributing entanglement over multiple nodes or active feedforward. We prove that secure quantum routing protects both the contents of the message and the identity of the sender and receiver, and creates only a negligible reduction in the network channel capacity.

**PACS:** 03.67.Hk, 03.67.Dd

---

[a)] Author to whom correspondence should be addressed. Email: edowaks@umd.edu.

Quantum networks enable unconditionally secure communication guaranteed by laws of physics[1]. The majority of quantum communication protocols to-date focus on securing messages being transmitted from one point to another[2-5]. When many quantum channels combine to form a complex network, however, the secrecy of messages transmitted between individual parties represents only one aspect of security. In many cases, knowledge of who is talking to whom in a network can reveal critical information. This realization lies at the heart of network traffic analysis, which focuses on using the flow of network data to extract useful information[6-9].

Network traffic analysis can reveal patterns in communication that can be useful to adversaries. For example, adversaries can target a desired person in the network by keying on patterns in their communication. Large scale mining of meta-data can also serve as the basis for developing predictive models that provide global information about the network. The fields of machine learning[10] and big data[11] are rapidly devising new methods to extract useful information from large dataset. At the same time, methods to protect networks against such attacks constitute an important security loophole.

Classical communication systems securitize network traffic flow based on packet reshuffling[12-15], shared randomness[16], or global broadcasting[17]. These protocols provide computational security, but not unconditional information theoretic security. More recently, Broadbent and Tapp showed that pairwise unconditionally secure channels can enable information theoretic anonymous communication that protects the identities of senders and receivers[18]. This protocol, however, works only for transmissions of classical bits and require large communication overhead that occupies a significant fraction of the network channel capacity. Following this work, several anonymous quantum protocols have been proposed with the ability to transmit quantum states with anonymity guaranteed by laws of physics[19-26]. However, these protocols all require establishing highly entangled Schrödinger cat states of the form $|\psi\rangle = (|00\cdots0\rangle + |11\cdots1\rangle)/\sqrt{2}$, where all the nodes of the network constitute a single entangled quantum system. Such complex entangled states can only be created using a long-lived quantum memory and some degree of distributed quantum computation capabilities, which is beyond current technological capabilities.

In this letter we demonstrate that a quantum network can protect the identity of senders and receivers from an external wiretapper without distributing complex Schrödinger cat states. We propose a protocol, which we denote secure quantum routing, that simultaneously protects the identity of a sender and receiver, as well as the content of the message from an eavesdropper that has access to all communication across the network. Implementation of the proposed protocol requires only efficient single photon sources and detectors along with linear optical elements, and does not require active feedforward. Therefore it can be implemented with currently available technological capabilities. Unlike anonymous communication, secure quantum routing does not protect the identity of the sender from the receiver. But it protects the identity of the sender and receiver from all third parties (including other nodes in the network), which protects the network from important security vulnerabilities such as traffic analysis.

Figure 1(a) illustrates our general network and eavesdropping model. We consider a network composed of an arbitrary number of point-to-point links. Each link is composed of an authenticated classical channel along with a quantum channel that can transmit quantum bits such as single photons. As with standard quantum key distribution (QKD), each node in the network constitutes a secure enclave where Eve cannot access or modify any of the equipment inside the node. The eavesdropper has access to all classical and quantum signals transmitted between the nodes for all time. We also assume that all the nodes can access a common clock to synchronize communication.

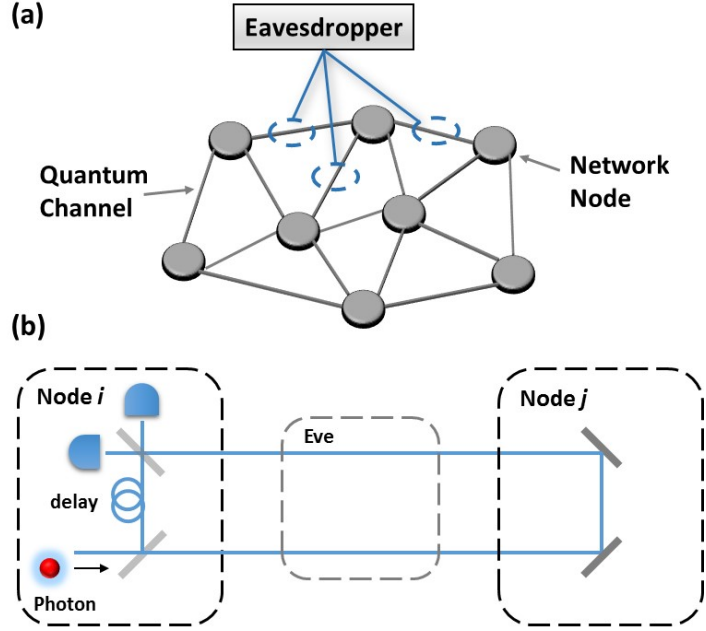

FIG. 1. (Color online) (a) Illustration of general network model. The network is composed of quantum nodes connected by quantum channels with arbitrary connectivity. The eavesdropper has access to all communications transmitted between nodes over the entire network. (b) Detailed schematic of a single quantum link between two nodes implementing a Type 3 communication. The structure is composed of an interferometer where half of the photon wavepacket remains protected in node $i$, while the other transmits back and forth from node $j$.

We represent a quantum transmission from node $i$ to node $j$ by the quantum state $|\psi\rangle = |i, j, n\rangle \otimes |x\rangle$, where $|i, j, n\rangle$ represents the spatial and temporal mode of the qubit propagating from node $i$ to $j$ on the $n$'th clock cycle, and $|x\rangle$ is the qubit state. Eve can acquire information about the sender and receiver of the message in two possible ways. First, she can monitor the transmitted messages $|x\rangle$ that may contain information about the sender and receiver's

identity. Second, she can monitor the spatio-temporal mode $|i,j,n\rangle$ which identifies that the qubit travels from the transmitter node *i* to the receiver node *j*. To protect against both types of attacks, we subdivide communication over the network into three categories, Type 1, Type 2 and Type 3 time slots. We allocate Type 1 time slots for regular network communication, Type 2 time slots to prevent an eavesdropper from monitoring $|x\rangle$, and Type 3 time slots to prevent an eavesdropper from monitoring the spatio-temporal mode $|i,j\rangle$. We will show that Type 2 and Type 3 time slots require only a small fraction of the total channel capacity.

The protocol proceeds as follows:

1. Prior to any network communication, all network nodes use standard QKD to exchange a secret message that tells which clock cycles are devoted to Type 2 and Type 3 communications, along with all other information required to implement Type 2 and Type 3 communications as described below.

2. In Type 1 time slots, if node *i* wants to transmit a qubit $|x\rangle$ to node *j*, it directly sends it over the quantum channel by preparing the state $|\psi\rangle = |i,j,n\rangle \otimes |x\rangle$. Physically, node *i* could prepare this qubit by generating a single photon and launching it at the appropriate time into a single mode fiber that connects to node *j*. After node *j* recognizes the receipt of the qubit, it sends a dummy qubit back to node *i* on the next clock cycle using the state $|\psi'\rangle = |j,i,n+1\rangle|y\rangle$ where $|y\rangle$ is a randomly prepared qubit.

3. In Type 2 time slots, node *i* randomly prepares a message qubit $|x\rangle$ in one of its two

eigenstates either using the basis $\{|0\rangle, |1\rangle\}$ or the basis $\{|+\rangle, |-\rangle\}$ where $|\pm\rangle = (|0\rangle \pm |1\rangle)/\sqrt{2}$. Node $i$ transmits the prepared state to a randomly selected node $j$ by preparing the state $|\psi\rangle = |i, j, n\rangle |x\rangle$. Node $j$ measures the message qubit in the same basis, and sends node $i$ a random qubit $|y\rangle$ on the next clock cycle. If node $j$ fails to receive a packet or detects more than one packets from node $i$ on the same clock cycle, it randomly generates a result.

4. In Type 3 time slots, node $i$ randomly prepares the qubit state in one of the following two states,

$$|\psi^{\pm}\rangle = \frac{1}{\sqrt{2}}(|i,i,n\rangle \pm |i, j, n\rangle)|y\rangle, \qquad (1)$$

where node $j$ is randomly selected in Step 1, and $|y\rangle$ is a randomly generated qubit. The state $|i,i,n\rangle$ represents the case where node $i$ generates a qubit and retains it within the node, rather than injecting into the network. The receiving node $j$ directly sends its component of the packet $|\psi^{\pm}\rangle$ back to node $i$ on the next clock cycle. Node $i$ then performs a measurement in the basis $|\pm\rangle_p = \frac{1}{\sqrt{2}}(|i,i,n+1\rangle \pm |j,i,n+1\rangle)$, ignoring the state of the randomly prepared qubit $|y\rangle$. If node $i$ fails to receive a packet or detects multiple packets within a clock cycle, they randomly generate a result. Figure 1(b) shows a physical implementation of this step, where the packet is composed of a single photon split into a superposition of two modes.

5. After a sufficient number of transmissions over the network, each node pair calculates the

Type 2 disturbance $D_2$, defined as the probability that node $j$ measures a different qubit state than the one sent by node $i$, and the Type 3 disturbance $D_3$, defined as the probability that node $i$ transmits state $|\psi^\pm\rangle$ and eventually detects the state $|\mp\rangle_p$. If either disturbance exceeds a threshold value, the network determines that a third party is eavesdropping and communication is discontinued.

To understand how the protocol protects the message, we first note that Eve does not know which clock cycles belong to Type 1, 2, or 3 time slots. Thus, if she monitors the quantum transmission she will inevitably measure some Type 2 and Type 3 time slots. If Eve attacks the content of the message during a Type 2 time slot, she will induce a disturbance in a completely analogous way to the BB84 protocol[5] because the communicating parties encode information in two non-orthogonal bases. If she instead performs a measurement on the spatio-temporal mode $|i,j,n\rangle$, she will induce disturbance during a Type 3 time slot by obtaining which-path information regarding whether the photon was transmitted to node $j$ or remained in the secure enclave of node $i$. As shown in Fig. 1(b), this which-path information destroys the interference at node $i$ once it receives the packet component back from node $j$.

We now prove that an eavesdropper cannot obtain information about the message and data traffic flow without inducing Type 2 and Type 3 disturbances respectively. We first define the most general eavesdropping model allowed by quantum physics. When node $i$ generates a quantum signal during a Type 1 time slot, the total state of the entire system, which includes Eve and the entire network, is given by $|\psi\rangle_{tot} = |\psi\rangle|R_n\rangle$, where $|\psi\rangle = |i,j,n\rangle|x\rangle$ is the state of the photonic packet and $|R_n\rangle$ is the state of the external environment on the *n*'th clock cycle that includes Eve's accessible Hilbert space, and the Hilbert space of all other nodes in the network which may or may not be exchanging quantum bits at the current clock cycle. This state can also depend on all communication history in the previous $n-1$ clock cycles. The only assumption is that the qubit produced by node *i* at the current clock cycle *n* is not initially entangled with the external environment. This assumption is valid because the qubit is produced locally inside node *i* which is a secure enclave that cannot be accessed by Eve.

Eve's measurement is composed of a unitary operation that entangles the state of the transmitted qubit with the external environment. We represent this operation mathematically as $|\phi_n\rangle = \mathbf{U}_n(|\psi\rangle|R_n\rangle)$, where $\mathbf{U}_n$ is a unitary operator that performs a general rotation in the Hilbert space spanned by $|\psi\rangle$ and $|R_n\rangle$. At each time slot Eve may select an optimal unitary operator to maximize the information she could obtain. Note that this attack does not limit Eve to a local attack on the channel between node *i* and *j*, because state $|R_n\rangle$ contains all past information available to Eve, as well as the present state of all other nodes in the network. Thus, this eavesdropping model represents a general coherent attack where Eve can exploit all past classical and quantum information, as well as the state of the entire network, to optimize her attack strategy.

To prove security, we first consider the constraints on Eve's unitary operator imposed by the requirement that she cannot induce a Type 2 disturbance. We expand the state using the spatio-temporal basis $|i,j,n\rangle$ which enables us to express the state of the system as

$$|\psi\rangle_{tot} = \sum_{k,l,m} |k,l,m\rangle \mathbf{U}_{ijn}^{klm} (|x\rangle|R_n\rangle), \qquad (2)$$

where $k$ and $l$ sum over all nodes in the network, $m$ sums over all clock cycles, and $\mathbf{U}_{ijn}^{klm} = \langle k,l,m|\mathbf{U}|i,j,n\rangle$ is an operator that acts on the subspace spanned by the message qubit $|x\rangle$ and $|R_n\rangle$. In Supplementary Section 1[27], we show that Eve cannot change the spatial or temporal mode of the photon, or injecting a photon of her own into the network. Doing so would result in photons being detected in the wrong node or at the wrong time which would induce Type 2 disturbance. In Eq. (2) these constraints are equivalent to the condition that $\mathbf{U}_{ijn}^{klm}$ is non-zero if and only if $i=k$, $j=l$, and $n=m$. Eq. (2) therefore simplifies to

$$|\psi\rangle_{tot} = |i,j,n\rangle \mathbf{U}_{ijn} (|x\rangle|R_n\rangle). \qquad (3)$$

where we use the simplified notation $\mathbf{U}_{ijn} = \mathbf{U}_{ijn}^{ijn}$.

In Supplementary Section 2[27], we prove that Eve cannot obtain information about the message without inducing a Type 2 disturbance. The proof proceeds in an analogous way to the BB84 protocol[28]. The reason for security is also the same as BB84. The sender encodes information in four non-orthogonal states, and any measurements on the system unavoidably induce a disturbance.

Now we examine communications during Type 3 time slots. These time slots serve to check if the eavesdropper is monitoring the propagation mode of the packets. We begin with the initial state that node $i$ prepares during a Type 3 time slot, denoted by $|\psi^\pm\rangle = \frac{1}{\sqrt{2}}(|i,i,n\rangle \pm |i,j,n\rangle)|y\rangle$. Eve applies unitary operations on clock cycle $n$ when node $i$ sends node $j$ the packet component $|i,j,n\rangle|y\rangle$, and on clock cycle $n+1$ when node $j$ returns the packet component back to node $i$. Once the wavepacket re-enters the secure enclave of node $i$, the state of the whole wavefunction

is

$$|\psi^{\pm}\rangle_{tot} = \frac{1}{\sqrt{2}}\left[|i,i,n+1\rangle\left(\mathbf{U}_{ii(n+1)}\mathbf{U}_{iin}|R_n\rangle\right) \pm |j,i,n+1\rangle\left(\mathbf{U}_{ji(n+1)}\mathbf{U}_{ijn}|R_n\rangle\right)\right]. \quad (4)$$

Because Eve cannot access the packet component within the enclave of node $i$, $\mathbf{U}_{iin} = \mathbf{V}_n$ where $\mathbf{V}_n$ is a unitary operator that is independent of node index $i$. To avoid Type 3 disturbance, the unitary operators must satisfy

$$\mathbf{U}_{ji(n+1)}\mathbf{U}_{ijn} = \mathbf{V}_{n+1}\mathbf{V}_n. \quad (5)$$

This condition must hold for all combinations of $i$ and $j$.

We now show that if Eve's unitary operation satisfies the constraints in Eq. (5), she attains no information during Type 1 time slots regarding which node is sending or receiving packets. In Type 1 communications, node $i$ transmits a packet at clock cycle $n$ given by the state $|i,j,n\rangle|x\rangle$, and node $j$ sends back a dummy packet $|j,i,n+1\rangle|y\rangle$ to node $i$ at clock cycle $n+1$. The state of the whole environment upon Eve's unitary operations is given by $|\psi\rangle_1 = \mathbf{U}_{ji(n+1)}\mathbf{U}_{ijn}|R_n\rangle$. If node $i$ does not transmit a packet at clock cycle $n$, the state of the whole environment after clock cycle $n+1$ is given by $|\psi\rangle_2 = \mathbf{V}_{n+1}\mathbf{V}_n|R_n\rangle$. According to Eq. (5), the state of the whole environment that includes Eve's own accessible system are identical and does not depend on the spatial mode of the transmitted packet. Thus, Eve's measurement results are independent whether node $i$ sent a qubit or whether node $j$ received a qubit. She cannot determine who is the sender or receiver of the packet, or even whether a packet was transmitted or not in the network.

The above proof shows that Eve cannot obtain information about the network data or traffic flow without inducing disturbance. In the asymptotic limit the network will detect Eve no matter how small of a disturbance she creates. But in a realistic network the length of communication is finite, so it is still possible for Eve to successfully attack the network without inducing disturbance. The network must assign sufficient number of Type 2 and Type 3 time slots to ensure that this probability is negligibly small, which results in a reduction in channel capacity. This reduction should be small in order for the protocol to be efficient. Furthermore, realistic networks will have a finite Type 2 and Type 3 disturbance even in the absence of eavesdropping. In the worst case, all disturbance must be attributed to eavesdropping and the network needs to estimate the amount of information that has been potentially leaked.

In order to analyze the efficiency and security of the protocol for a realistic network in the presence of noise and loss, we consider a restricted set of eavesdropping attacks based on an intercept and re-send strategy. In this attack Eve intercepts a packet and measures either the propagation mode or the content of the message (or both). She then relays the measured packet to the intended receiver. For each packet where Eve measures the propagation mode, she learns complete information about the sender and receiver. We note that although this attack is not the most general attack, it still gives Eve complete access to the entire network simultaneously, which is extremely challenging using current technological capabilities.

We first investigate the efficiency of the protocol. During Type 2 and Type 3 time slots a node pair cannot engage in normal network communication, which reduces the overall channel capacity. Furthermore, prior to the protocol the network must exchange additional information such as which clock cycles constitute Type 2 and Type 3 communications and who are the receivers and senders, and after the protocol the network must exchange measurement results in each Type 2 and Type 3 communication slots to calculate disturbance. These communications further degrade the total channel capacity. We define the overhead $H$ as the average number of bits transmitted per node pair prior to network communication, during Type 2 and Type 3 time slots, and after communication to calculate disturbance. We calculate $H$ for a network with $K$ communication time slots. In Supplementary Section 3 we show that $H = O(\log K)^{27}$, which scales logarithmically with the number of communication time slots, demonstrating that the protocol is efficient.

We next consider the security of the protocol in the presence of finite Type 2 and Type 3 disturbance. We focus on the case where Eve aims to obtain the information about the network traffic flow, not the message. The analysis of the message security has already been extensively studied for QKD protocols[29], and these results are applicable to our protocol as well. If Eve intercepts a fraction $\eta$ of the total communication clock cycles, she learns the identity of the sender and receiver for a fraction $\eta$ of the transmitted bits, but she also induces a Type 3 disturbance of $D = \eta/2$. This information disturbance tradeoff enables us to calculate the maximum amount of information leaked to Eve given a measured disturbance.

A realistic network will have a baseline disturbance level due to losses and measurement error. The disturbance for the link between node $i$ and node $j$ is $D_{ij} = \gamma_{ij} T_{ij} T_{ji} + (1 - T_{ij} T_{ji})/2$, where $\gamma_{ij}$ is the baseline error probability of the interference measurement due to finite visibility of the interferometer, and $T_{ij}$ is the probability that a photon successfully transmits from node $i$ to node $j$. In the following analysis, we assume that $T_{ij}$ and $\gamma_{ij}$ are identical for all channels in the network so that we can denote $D_{ij}$, $T_{ij}$ and $\gamma_{ij}$ as $D$, $T$ and $\gamma$ respectively.

To achieve error-free communication, a sender must transmit a message $M = [A, E]$, where $A$ is the actual message and $E$ is additional information required to perform error correction. The total error probability of the communication is given by $e = \mu T + (1-T)/2$, where $\mu$ is the baseline error probability for the communication channel. An error correction algorithm operating at the Shannon limit requires the length of $M$ to be $L(1 + h(e))$, where $L$ is the length of $A$ and $h(e) = -e\log(e) - (1-e)\log(1-e)$. Thus, the fraction of network traffic flow information obtained by an eavesdropper is given by

$$g = \max\{1, 2D(1 + h(e))\}. \tag{6}$$

Figure 2 plots the fraction of network traffic flow information obtained by an eavesdropper as a function of channel loss (in dB), where we set $\gamma = 0.01$ and $\mu = 0.01$. The fraction of network traffic flow information that Eve could obtain increases monotonically with respect to channel losses, because losses create Type 3 disturbance even in the absence of eavesdropping. The communication protocol can withstand 1.9 dB loss before the whole network traffic flow information is revealed to an eavesdropper.

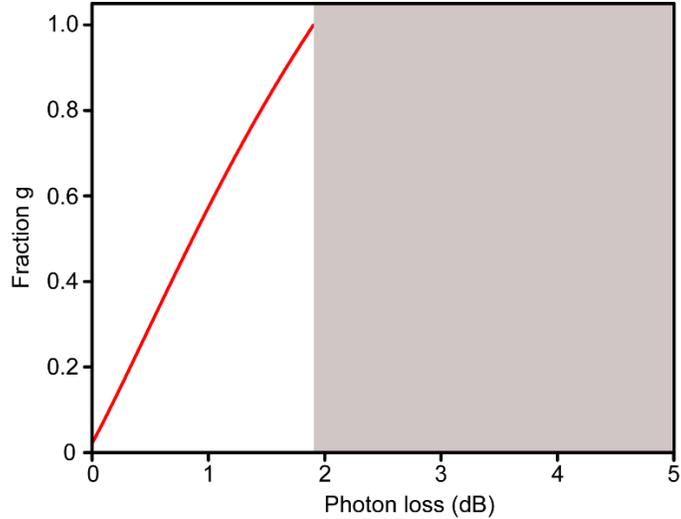

FIG. 2. (Color online) Fraction of network traffic flow information obtained by an eavesdropper as a function of photon loss (in dB). The grey region represents the regime where traffic flow is completely revealed to an eavesdropper.

In summary, we demonstrate that a quantum network capable of exchanging qubits can protect the network traffic flow information from an external eavesdropper without distributing highly entangled states of many nodes. This information is not limited to the identities of the sender and receiver of each transmitted packet. It also protects global properties of the network. For example, the eavesdropper cannot distinguish a silent network from one that is transmitting many packets. The dominant mechanism for degraded security is channel loss. An important open question is whether better algorithms exist that can handle loss more efficiently. Another important question is how to incorporate classical reconciliation into secure quantum routing in order to more efficiently handle the security of the message. In general, our protocol provides compelling approach to extend conventional point to point quantum communication to more complex network settings to attain quantum advantages over classical networks.

The authors would like to thank T. Paul, A. Agrawal, and T. Chapuran from Vencore Labs, Inc. for helpful discussions. The authors would also like to acknowledge support from the Laboratory for Telecommunication Sciences, the DARPA QUINESS program (grant number W31P4Q1410003), the Physics Frontier Center at the Joint Quantum Institute, and the National Science Foundation (grant number PHYS. 1415458).

**Figure Captions**

FIG. 1 (Color online) (a) Illustration of general network model. The network is composed of quantum nodes connected by quantum channels with arbitrary connectivity. The eavesdropper has access to all communications transmitted between nodes over the entire network. (b) Detailed schematic of a single quantum link between two nodes implementing a Type 3 communication. The structure is composed of an interferometer where half of the photon wavepacket remains protected in node *i*, while the other transmits back and forth from node *j*.

FIG. 2. (Color online) Fraction of network traffic flow information obtained by an eavesdropper as a function of photon loss (in dB). The grey region represents the regime where traffic flow is completely revealed to an eavesdropper.

# Supplementary Materials

# Secure quantum routing

Shuo Sun and Edo Waks

## 1. Security proof that Eve cannot change the spatial or temporal mode of the photon

We show that Eve cannot change the spatial or temporal mode of the photon, or injecting a photon of her own into the network. We note that in order to avoid inducing Type 2 disturbance, Eve must ensure that node $j$ receives the packet sent by node $i$. If Eve diverts the packet to another node or to the same node at a different clock cycle, node $j$ will fail to receive a qubit at the correct clock cycle which induces Type 2 disturbance with a probability of 0.5. Using our notation, this restriction means that Eve's unitary operator cannot transform mode $|i,j,n\rangle$ to another mode $|k,l,m\rangle$ or retain it within her own system. Furthermore, Eve cannot inject additional photons into the mode $|i,j,n\rangle$ because node $j$ will detect multiple packets, which also induces Type 2 disturbance with a probability of $0.5$[1,2]. Finally, Eve cannot exchange the packet in the mode $|i,j,n\rangle$ with a different one from another part of the network or from her own system. This action will clearly induce Type 2 disturbance because the messages transmitted in other packets in the network have no correlation with the qubit component sent from node $i$ to $j$ during a Type 2 time slot. In Eq. (2) of the main manuscript, these constraints are equivalent to the condition that $\mathbf{U}_{ijn}^{klm}$ is non-zero if and only if $i=k$, $j=l$, and $n=m$. Eq. (2) therefore simplifies to Eq. (3) in the main manuscript.

## 2. Security proof that Eve cannot obtain information about transmitted message

We show that Eve cannot obtain information about the message without inducing a Type 2 disturbance. The proof proceeds in an analogous way to the BB84 protocol[3]. During a Type 2 time slot node $i$ randomly prepares the qubit state $|x\rangle$ either in the states $\{|0\rangle,|1\rangle\}$ or $\{|+\rangle,|-\rangle\}$. Using the first basis, we can expand Eve's unitary operator as

$$\mathbf{U}_{ijn}\left(|0\rangle|R_n\rangle\right) = |0\rangle \mathbf{U}_{ijn}^{00}|R_n\rangle + |1\rangle \mathbf{U}_{ijn}^{10}|R_n\rangle, \tag{S1}$$

$$\mathbf{U}_{ijn}\left(|1\rangle|R_n\rangle\right) = |0\rangle \mathbf{U}_{ijn}^{01}|R_n\rangle + |1\rangle \mathbf{U}_{ijn}^{11}|R_n\rangle, \tag{S2}$$

where $\mathbf{U}_{ijn}^{ab} = \langle a|\mathbf{U}_{ijn}|b\rangle$. Eve cannot flip the state of the message qubit since this operation will lead to Type 2 disturbance, which means $\mathbf{U}_{ijn}^{01} = \mathbf{U}_{ijn}^{10} = \mathbf{0}$. If node $i$ instead uses the second basis to prepare the qubit $|x\rangle$, the final state becomes

$$\mathbf{U}_{ijn}\left(|\pm\rangle|R_n\rangle\right) = \frac{1}{\sqrt{2}}\left[|+\rangle\left(\mathbf{U}_{ijn}^{00} \pm \mathbf{U}_{ijn}^{11}\right) + |-\rangle\left(\mathbf{U}_{ijn}^{00} \mp \mathbf{U}_{ijn}^{11}\right)\right]|R_n\rangle. \tag{S3}$$

To avoid Type 2 disturbance in this basis requires $\mathbf{U}_{ijn}^{00} = \mathbf{U}_{ijn}^{11}$. Thus, the final state of the system is $|\psi\rangle_{tot} = |i,j,n\rangle|x\rangle\left(\mathbf{U}_{ijn}|R_n\rangle\right)$ where the unitary operator $\mathbf{U}_{ijn}$ does not depend on the state of the message qubit. Because Eve's probe is completely disentangled from the message qubit, her measurement results are uncorrelated with the qubit state and she gains no information about the message. We note that in principle $\mathbf{U}_{ijn}$ could apply a constant qubit-independent rotation on the message Hilbert space, but such a rotation merely constitutes a change in Eve's measurement basis. Thus, we may assume without loss of generality that the unitary operators $\mathbf{U}_{ijn}$ only operate on $|R_n\rangle$ not on $|x\rangle$.

### 3. Calculation of communication overhead

Let $P(\eta)$ be the probability that Eve intercepts a fraction $\eta$ of the total communication slots between any two nodes without causing any disturbance. We prove that, for any constant $\varepsilon$ such that $0 < \varepsilon < 1$, and any constant $\eta_{max}$ such that $0 < \eta_{max} < 1$, there exist constant numbers $g_0$ and $g_1$ which are independent of $K$, such that if $H = g_1 \log(K) + g_0$ then $\lim_{K \to \infty} P(\eta) < \varepsilon$ is satisfied for any $\eta > \eta_{max}$. This statement shows that, for an arbitrarily small $\eta_{max}$ and $\varepsilon$, we can always find constant numbers $g_0$ and $g_1$ such that the probability that Eve intercepts more than a fraction $\eta_{max}$ of the total communication slots without inducing disturbance is smaller than $\varepsilon$. Thus, the required overhead scales only logarithmically with the number of total communication slots.

To prove this statement, we write $H$ as $H = H_1 + H_2 + H_3 + H_4$, where $H_1$ is the number of exchanged bits between two nodes in the first step of the protocol, $H_2$ and $H_3$ is the number of exchanged bits (qubits) between two nodes during Type 2 and Type 3 time slots respectively, and $H_4$ is the number of exchanged bits between two nodes in the last step of the protocol for calculating the disturbance.

**Theorem 1.** $H_1 = H_2 \cdot (\log K + 1) + H_3 \cdot \log K$.

*Proof.* In the first step of the protocol, each node in the network must know when to send or receive a Type 2 and Type 3 packet from another node. Since it takes $\log K$ bits to specify one time slot within $K$ time slots, this step takes $(H_2 + H_3) \cdot \log K$ bits. For each Type 2 packet, the sender and receiver also needs to know the basis to prepare and measure the message qubit.

Since it takes 1 bit to specify the basis for each Type 2 packet, this step takes a total number of $H_2$ bits. We therefore obtain that $H_1 = H_2 \cdot (\log K + 1) + H_3 \cdot \log K$.

**Theorem 2.** $H_4 = H_2 + H_3$.

*Proof.* In the last step of the protocol, each node pair in the network exchanges the measured result for each Type 2 and Type 3 packet. Since it takes 1 bit to specify the outcome of the measurement, this step takes $H_2 + H_3$ bits. We therefore obtain that $H_4 = H_2 + H_3$.

The following Lemma is necessary for the proof of **Theorem 3**.

**Lemma 1.** Suppose Eve intercepts a fraction $\eta$ of the total communication slots between node $i$ and node $j$ and measures the propagation mode of each intercepted packet. The probability $P(\eta)$ that Eve does not induce any Type 3 disturbance satisfies $P(\eta) < \left( \dfrac{K - H_3}{K} + \dfrac{H_3}{2(1-\eta)K} \right)^{\eta K}$.

*Proof.* We consider the case where $\eta < H_3/K$ and $\eta \geq H_3/K$ separately. In the case where $\eta < H_3/K$, We can write $P$ as

$$P = \sum_{m=0}^{\eta K} P_m, \tag{S4}$$

where $P_m$ represents the probability that Eve does not induce any Type 3 disturbance even when she intercepts $m$ Type 3 time slots. The expression of $P_m$ is given by

$$P_m = \frac{\binom{H_3}{m} \times \binom{H_0}{\eta K - m}}{\binom{K}{\eta K}} \times \left( \frac{1}{2} \right)^m. \tag{S5}$$

where $H_0 = K - H_3$. We can rewrite $P_m$ as

$$P_m = \binom{\eta K}{m} \times \frac{H_3 \cdot (H_3 - 1) \cdots (H_3 - m + 1) \times H_0 \cdot (H_0 - 1) \cdots (H_0 - \eta K + m + 1)}{K \cdot (K-1) \cdots (K - \eta K + 1)} \times \left(\frac{1}{2}\right)^m. \tag{S6}$$

Using the identity $\dfrac{x-a}{y-a} < \dfrac{x}{y}$ for $y > x > a > 0$, we obtain that

$$P_m < \binom{\eta K}{m} \times \left(\frac{H_0}{K}\right)^{\eta K - m} \times \left(\frac{H_3}{2(K - \eta K)}\right)^m. \tag{S7}$$

Substituting Eq. (S7) into Eq. (S4), we obtain that

$$P < \left(\frac{K - H_3}{K} + \frac{H_3}{2(1-\eta)K}\right)^{\eta K}. \tag{S8}$$

In the case where $\eta \geq H_3/K$, we can write $P$ as

$$P = \sum_{m=0}^{H_3} P_m \leq \sum_{m=0}^{\eta K} P_m, \tag{S9}$$

where $P_m$ still satisfies Eq. (S7). Substituting Eq. (S7) into Eq. (S9), we obtain Eq. (S8) again, which completes the proof.

**Lemma 2.** If $P(\eta_1, H_3) < a$, then $P(\eta_2, H_3) < a$ as long as $0 < \eta_1 \leq \eta_2 < 1$.

*Proof.* For the same number of Type 3 time slots, the probability that Eve does not get detected would drop if Eve intercepts more packets. Therefore we have $P(\eta_2, H_3) \leq P(\eta_1, H_3) < a$ for $\eta_2 \geq \eta_1$.

**Theorem 3.** Let $P(\eta)$ be the probability that Eve intercepts a fraction $\eta$ of the total communication slots without causing any disturbance. For any constant $\varepsilon$ such that $0 < \varepsilon < 1$, and any constant $\eta_{max}$ such that $0 < \eta_{max} < 1$, there exists a constant number $\alpha$ which is independent of $K$, such that if $H_3 = \alpha$ then $\lim_{K \to \infty} P(\eta) < \varepsilon$ is satisfied for any $\eta > \eta_{max}$.

*Proof.* Using **Lemma 1**, we obtain that

$$\lim_{K \to \infty} P(\eta_{max}) < \lim_{K \to \infty} \left( \frac{K - H_3}{K} + \frac{H_3}{2(1-\eta_{max})K} \right)^{\eta K} = \exp\left( -\frac{\eta_{max} H_3 (1 - 2\eta_{max})}{2(1-\eta_{max})} \right). \quad (S10)$$

For any $\eta$ that satisfies $\eta > \eta_{max}$, using **Lemma 2**, we have

$$\lim_{K \to \infty} P(\eta) < \lim_{K \to \infty} P(\eta_{max}) \leq \lim_{\eta_{max} \to 0, K \to \infty} P(\eta_{max}) \quad (S11)$$

Substituting Eq. (S10) into Eq. (S11), we obtain that

$$\lim_{K \to \infty} P(\eta) < \exp\left( -\frac{1}{2} \eta_{max} H_3 \right). \quad (S12)$$

We let $\alpha = \frac{2}{\eta_{max}} \ln\left(\frac{1}{\varepsilon}\right)$, which is independent of $K$. It is straightforward to verify that when $H_3 = \alpha$, then $\lim_{K \to \infty} P(\eta) < \varepsilon$.

**Theorem 4.** Let $P(\eta)$ be the probability that Eve measures the message qubit state for a fraction $\eta$ of the total transmitted packets without causing any disturbance. For any constant $\varepsilon$ such that $0 < \varepsilon < 1$, and any constant $\eta_{max}$ such that $0 < \eta_{max} < 1$, there exists a constant number $\beta$ which is independent of $K$, such that if $H_2 = \beta$ then $\lim_{K \to \infty} P(\eta) < \varepsilon$ is satisfied for any $\eta > \eta_{max}$.

*Proof.* This proof follows the same procedure as the proof for **Theorem 3**.

Finally, by combining **Theorem 1-4** and the definition of $H$, we obtain that $H = (\alpha + \beta) \log K + (\alpha + 2\beta)$. Setting $g_0 = \alpha + 2\beta$ and $g_1 = \alpha + \beta$ completes the proof.